# Instant ghost imaging: improving robustness for ghost imaging subject to optical background noise

### ZHE YANG, WEI-XING ZHANG, MA-CHI ZHANG, DONG RUAN, AND JUN-LIN LI*

*State Key Laboratory of Low-dimensional Quantum Physics and Department of Physics, Tsinghua University, Beijing 100084, China*
*\*center@mail.tsinghua.edu.cn*

**Abstract:** Ghost imaging (GI) is an imaging technique that uses the second-order correlation between two light beams to obtain the image of an object. However, standard GI is affected by optical background noise, which reduces its practical use. We investigated the robustness of an *instant ghost imaging* (IGI) algorithm against optical background noise and compare it with the conventional GI algorithm. Our results show that IGI is extremely resistant to spatiotemporally varying optical background noise that can change over a large range. When the noise is large in relation to the signal, IGI will still perform well in conditions that prevent the conventional GI algorithm from generating an image because IGI uses signal differences for imaging. Signal differences are intrinsically resistant to common noise modes, so the IGI algorithm is strongly robust against noise. This research is of great significance for the practical application of GI.

## 1. Introduction

Ghost imaging (GI) is an imaging technique that generates the image of an object by calculating the second-order correlation function between two light beams [1-6]. Ghost imaging has been widely researched in recent years [7-16]; it has important applications in many fields such as cryptography [17,18], lidar [19,20], medical imaging [21,22], micro object imaging [23,24], three-dimensional imaging [25-28] and single-pixel imaging [29-33]. In many practical scenes, GI is subject to interference from the transmission medium and from optical background noise. Many studies have demonstrated that, in the first case, GI is more robust against disturbance from medium turbulence and medium scattering than traditional imaging techniques [34-38]. In the latter case, interference from optical background noise has been shown to decrease the quality of a lidar image [39]. The imaging capability of ghost imaging is reduced, or possibly even disabled, when variation in the amplitude of noise light exceeds the intensity of the thermal light. This problem is an obstacle to the practical application of GI.

[40] prototyped the chip-based application of an instant ghost imaging (IGI) algorithm. IGI uses the difference between the signals of two consecutive measurements to generate an image of the object [41-43]. This signal difference approach suggests that IGI can greatly increase GI robustness in respect of optical background noise because signal difference has an intrinsic denoising capability [44,45].

In this paper, to produce optical background noise of given frequency and intensity, we used a controlled LED. By comparing the imaging performance between the IGI algorithm and the GI algorithm under the same levels of optical background noise, we found that the IGI algorithm has great robustness against markedly varying degrees of optical background noise. Besides, the IGI algorithm could still produce a high-quality image when the GI algorithm failed, signifying its superiority.

## 2. Method

### 2.1 Experimental configuration

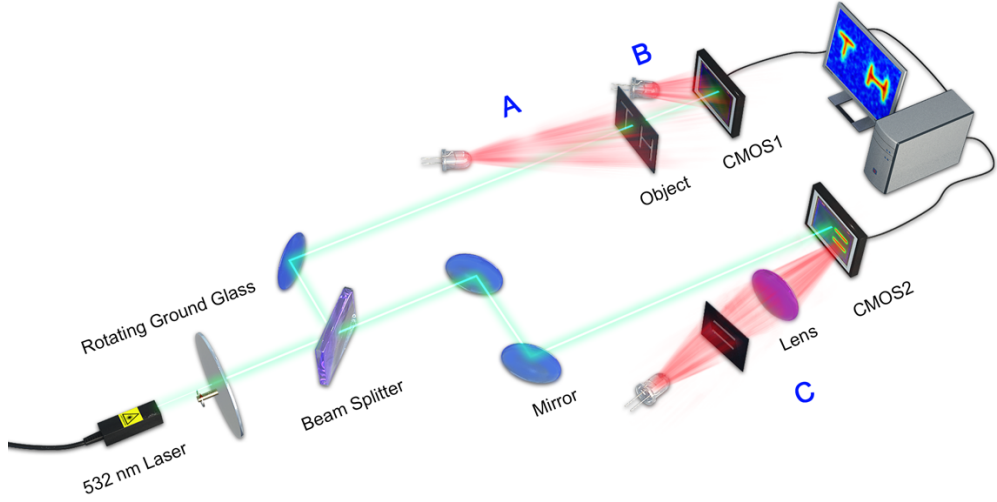

**Fig. 1.** The experimental configuration; components are described in the text.

The experimental configuration is shown in Fig. 1. A 532 nm laser beam passes through the rotating ground glass screen to produce pseudo-thermal light. A beam splitter divides the light beam into two spatially correlated beams, the test beam and the reference beam. A mask bearing the letters TH, the object, is placed in the path of the test beam, and CMOS1 is the bucket detector. The reference beam is directed to CMOS2, which detects the reference light. The distance from the ground glass to the object is 300 mm, the same as the distance from the ground glass to CMOS2; the object is close to CMOS1.

The optical background noise is generated by an FPGA driving a white LED. The noise source is placed at one of positions A, B, and C so that noise is introduced between the source and the object, between the object and the bucket detector CMOS1, or between the source and the reference signal detector CMOS2. The noise is introduced to enable us to investigate the robustness of the IGI and GI algorithms against optical background noise. The detection rate of each CMOS is 25 measurements per second, and the total number of measurements N is 20000.

### 2.2 Imaging reconstruction algorithms

The image of the object is reconstructed by calculating the second-order correlation between the two beams. The conventional background subtraction GI algorithm is [3,28]

$$G(x) = \left\langle [S - \langle S \rangle][I(x) - \langle I(x) \rangle] \right\rangle \tag{1}$$

where $S$ is the value of the bucket detector, which is calculated by summing the intensities of all pixels on CMOS1; $I(x)$ is the intensity of the reference beam at coordinate $x$; and the ensemble average is $\langle \cdot \rangle = (1/N) \sum_{i=1}^{N} (\cdot)$.

The IGI algorithm is

$$G^{IGI}(x) = \frac{1}{2N} \sum_{n=1}^{N} [S_{n+1} - S_n][I_{n+1}(x) - I_n(x)] \tag{2}$$

where $S_{n+1} - S_n$ and $I_{n+1}(x) - I_n(x)$ are the temporal difference signals between two consecutive measurements at the bucket detector and the reference detector. The result given by IGI is the same as that given by GI; the detailed proof is given in [40].

## 3. Results

### 3.1 Study of position A

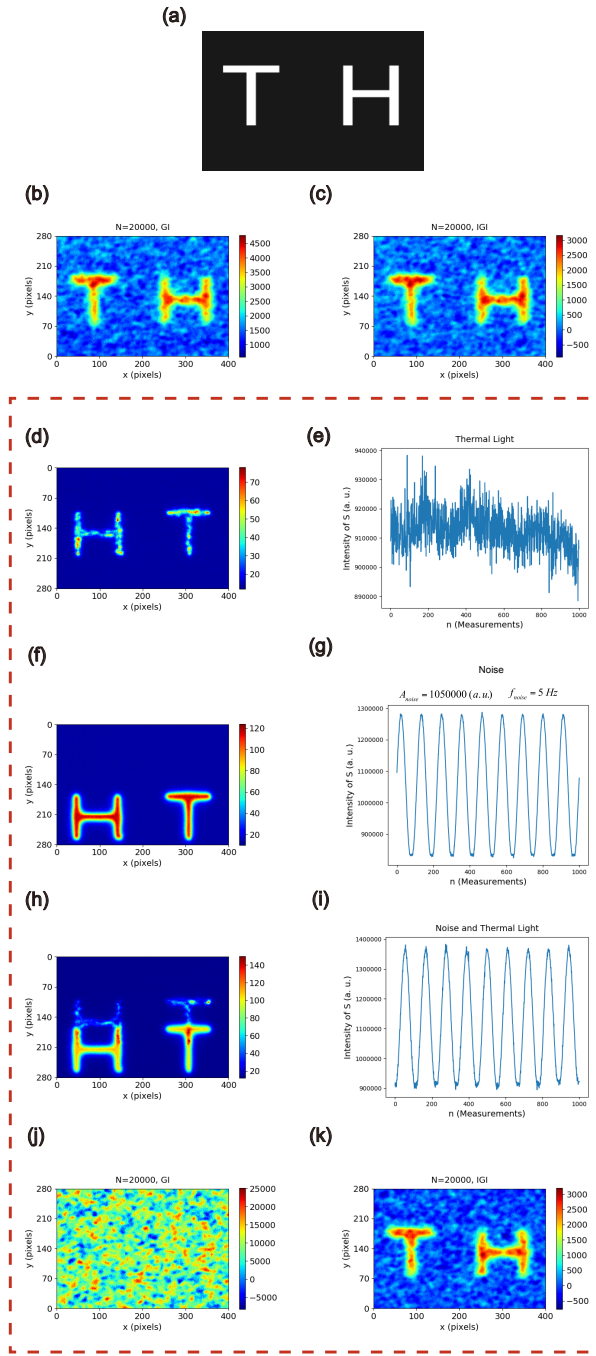

**Fig. 2.** Noise resistance for position A; individual images are explained in the text.

We verified that both the GI algorithm and the IGI algorithm could reconstruct the image of the object when there is no optical background noise. Fig. 2(a) shows the object. Figs. 2(b) and 2(c) are the images of the object reconstructed respectively by the GI algorithm and the IGI algorithm. We found that the two algorithms produce almost identical quality images when there is no optical background noise.

We studied the robustness of the system when the optical noise source was at position A. We placed the LED in front of the object, and the FPGA produced optical background noise of varying intensity. Fig. 2(d) shows an image obtained on CMOS1 when there was no optical background noise (i.e., with only thermal light illuminating the object); Fig. 2(e) is the corresponding value of $S$ at the bucket detector (for the convenience of display, only the first 1000 measurements are shown). Fig. 2(f) is a photograph obtained on CMOS1 when the object is illuminated only by optical background noise; the corresponding $S$ curve is shown in Fig. 2(g). The amplitude of the added noise with trigonometric function distribution is 1050000 (a. u.) and the frequency is 5 Hz. Fig. 2(h) is a photograph obtained on CMOS1 when the object is illuminated by both optical background noise and thermal light; the corresponding $S$ curve is shown in Fig. 2(i). Figs. 2(j) and 2(k) are the images of the object formed by the GI algorithm and the IGI algorithm.

It can be seen that when the test beam was subject to interference from optical background noise, the GI algorithm was completely unable to generate an image, but the IGI algorithm was almost unaffected by the noise. In position A, IGI is extremely robust against greatly changing levels of optical background noise; therefore, its imaging capability is much greater than that of the traditional GI algorithm.

*3.2 Study of position B*

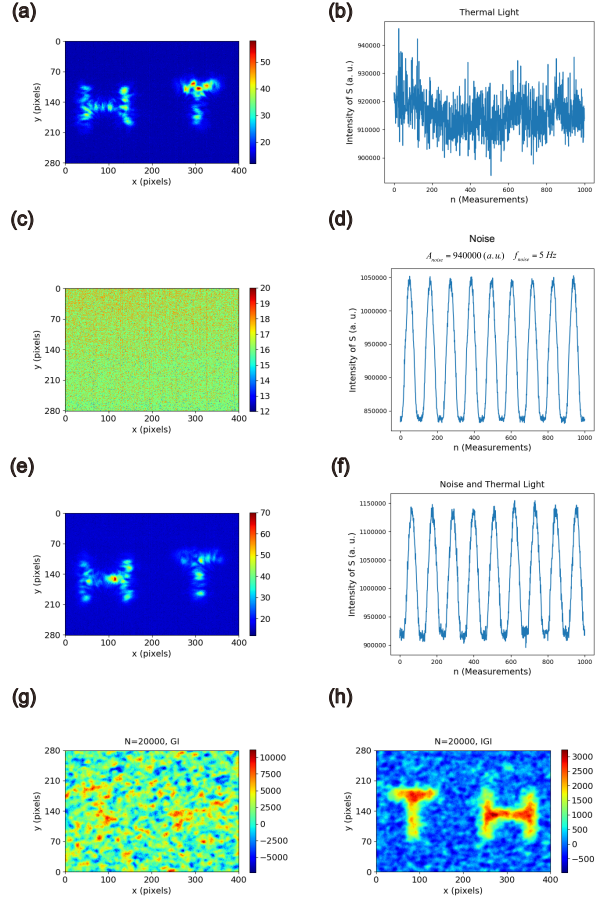

**Fig. 3.** Noise resistance study for position B; individual images are explained in the text.

We investigated the robustness of the GI and IGI algorithms when the optical background noise source was at position B. We placed the LED between the object and CMOS1. Light intensity was varied by the FPGA.

Fig. 3(a) shows a photo obtained by CMOS1 when the object was illuminated only by thermal light; Fig. 3(b) shows the $S$ curve of the bucket detector when there was no background noise. Fig. 3(c) shows a photo obtained by CMOS1 when the object was illuminated only by optical background noise; Fig. 3(d) shows the corresponding values of the $S$ curve at the bucket detector CMOS2. The amplitude of the added noise with trigonometric function distribution is 940000 (a. u.) and the frequency is 5 Hz. Fig. 3(e) shows a photo captured by CMOS1 when the object was illuminated by both thermal light and optical background noise; Fig. 3(f) shows the corresponding values of the $S$ curve at the bucket detector.

Figs. 3(g) and 3(h) are the images of the object generated respectively by the GI and IGI algorithms. We can see that when the collector CMOS1 is subject to time-variant optical background noise, GI is completely unable to image, but IGI is almost unaffected. When background noise is introduced at position B, IGI is extremely robust against widely varying levels of optical background noise; therefore, its imaging capability is much greater than that of the traditional GI algorithm.

*3.3 Study of position C*

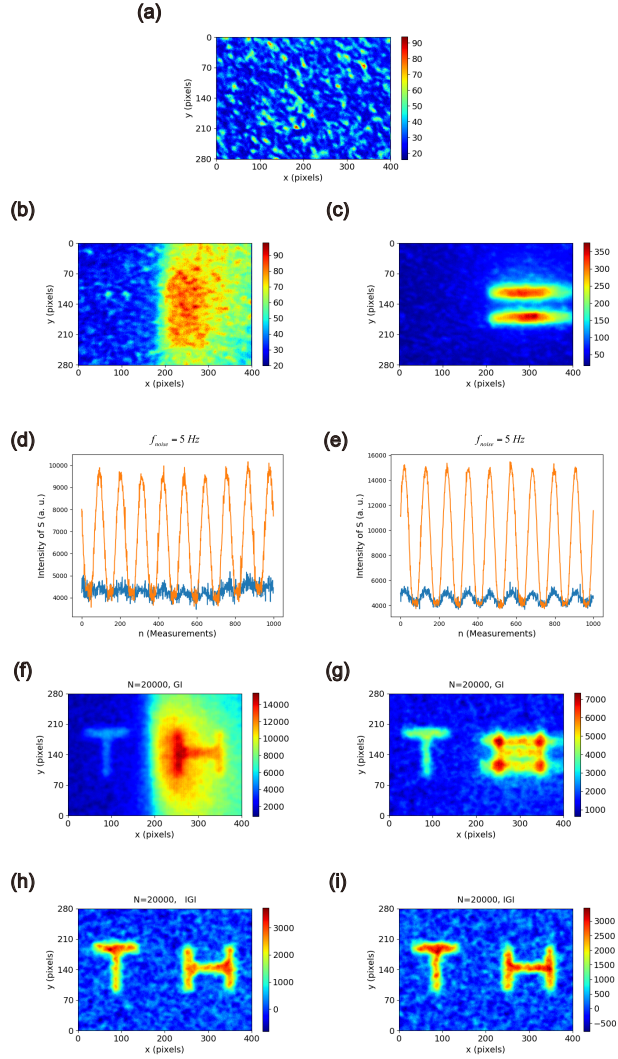

**Fig. 4.** Noise resistance study for position C; individual images are explained in the text.

We investigated the robustness of both algorithms when the optical background noise source was at position C. We placed the LED between the object and CMOS2 so that the optical background noise source illuminated CMOS2. The intensity of the light was varied by the FPGA. We compared the imaging capability of the GI and IGI algorithms for varying levels of background noise.

Fig. 4(a) is a speckle image captured by CMOS2 when the CMOS2 was illuminated only by the reference beam (i.e., there was no introduced optical background noise). Fig. 4(b) and Fig. 4(c) are the speckle images captured by CMOS2 when it was illuminated by both the thermal light and the optical background noise source at position C. In Fig. 4(b), optical background noise illuminates the right half of CMOS2; in Fig. 4(c), double-slit optical background noise illuminates the right half of CMOS2. The frequency of the added trigonometric function noise is 5 Hz. In Figs. 4(d) and 4(e), the blue curve shows reference light fluctuation for the background noise-free portion (left-hand side of CMOS2; the summation of the 100th column); the orange curve shows reference light fluctuation for the background noise-illuminated portion (right-hand side of CMOS2; the summation of the 300th column). Fig. 4(f) and Fig. 4(g) are images of objects generated by the GI algorithm; Fig. 4(h) and Fig. 4(i) are images of objects generated by the IGI algorithm.

It can be seen that interference due to the spatiotemporally distributed optical background noise had a great influence on GI because the image quality is significantly degraded. However, it had little effect on IGI: the image quality is almost unchanged.

## 4. Discussion

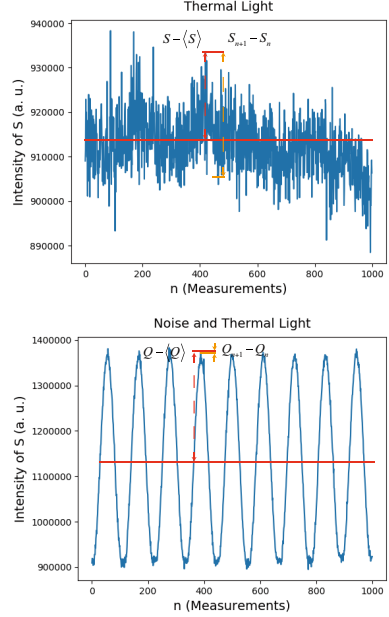

**Fig. 5.** Analysis of the IGI algorithm and the GI algorithm when there is optical background noise.

In this section, we discuss the reason for the IGI algorithm being more resistant to background noise than the conventional GI algorithm.

First, we analyze the case where the optical background noise is added to the test light path (position A and position B). We assume that the background noise intensity is $Q$. The GI algorithm is written as

$$G(x) = \left\langle [S - \langle S \rangle + Q - \langle Q \rangle][I(x) - \langle I(x) \rangle]\right\rangle \tag{3}$$

Fig. 5 shows that $Q - \langle Q \rangle$ is greater than $S - \langle S \rangle$, which is indicated by the red dash line. In this case, light fluctuations in the test beam are annihilated by the noise fluctuations. Thus, the GI algorithm cannot obtain the image of the object.

For the IGI algorithm subjected to optical background noise,

$$G^{IGI}(x) = \frac{1}{2N}\sum_{n=1}^{N}[S_{n+1} - S_n + Q_{n+1} - Q_n][I_{n+1}(x) - I_n(x)] \tag{4}$$

Given the background noise configuration shown in Fig. 5, $Q$ can be regarded as a slowly varying signal related to $S$, since the frequency of change in $Q$ is much smaller than that in $S$, which means that $Q_{n+1} - Q_n$ is approximately zero; it is also much smaller than $S_{n+1} - S_n$, which is indicated as the orange dash line. That is, the noise fluctuations are much smaller than the fluctuations in the test beam, and thus the effect on the quality of the IGI image is almost negligible. Therefore, the IGI algorithm is more resistant to time-variant optical background noise than the GI algorithm.

We now discuss the case where CMOS2 is subjected to background noise from position C. The background noise is represented by $Q(x)$. The GI algorithm is written as

$$G(x) = \left\langle [S - \langle S \rangle][I(x) - \langle I(x) \rangle + Q(x) - \langle Q(x) \rangle] \right\rangle \qquad (5)$$

For points at which $Q(x)$ is nonzero, when $Q - \langle Q \rangle$ is approximately equal to or greater than $S - \langle S \rangle$, the background noise will have a large effect on the result at coordinate $x$, leading to a significant degradation in GI image quality.

For the IGI algorithm, in this case, we write

$$G^{IGI}(x) = \frac{1}{2N} \sum_{n=1}^{N} [S_{n+1} - S_n][I_{n+1}(x) - I_n(x) + Q_{n+1}(x) - Q_n(x)] \qquad (6)$$

Since the frequency of change in $Q(x)$ is much less than the frequency of change in $I(x)$, $Q(x)$ can be regarded as a slowly varying signal related to $I(x)$; thus, $Q_{n+1}(x) - Q_n(x)$ is approximately zero and is much smaller than $I_{n+1}(x) - I_n(x)$. That is, the noise is much less than the fluctuations in the thermal light of the reference beam. Therefore, the results given by IGI are only slightly affected by noise. Thus, the IGI algorithm is more resistant to spatiotemporally varying optical background noise than the GI algorithm.

In the flowing, will discuss what type of noise can be or cannot be rejected by the IGI algorithm. The essence of the IGI algorithm is to use the difference between consecutive temporal signals to generate the image of the object. We will specify the condition under which the IGI algorithm works well.

From the time domain perspective, the IGI works well as long as the difference value of the optical background noise is far less than the difference result of the signal,

$$(Q_{n+1} - Q_n) << (S_{n+1} - S_n)$$
$$[Q_{n+1}(x) - Q_n(x)] << [I_{n+1}(x) - I_n(x)] \qquad (7)$$

From this point, as long as the above conditions are met, the IGI algorithm can handle any type of noise.

However, the anti-noise capability of the IGI algorithm is limited. If Eq. (7) is not satisfied with trigonometric function noise, Poisson noise, and Gaussian white noise, the imaging quality reconstructed by the IGI algorithm will decrease and even be destroyed.

## 5. Summary and Conclusion

We investigated the comparative robustness of the IGI algorithm and a GI algorithm against optical background noise. We introduced the noise separately at each of three different locations in the GI equipment. The IGI algorithm was significantly more robust against optical background noise than the conventional GI algorithm. This was because the IGI algorithm utilizes the difference between two consecutive signal measurements to generate the image of the object. The signal difference effectively eliminates the noise component of the signal, which improves the ability of IGI to withstand interference from optical background noise.

The IGI algorithm offers additional advantages because of its robustness against optical background noise: it is more easily generalizable (it could be used for IGI radar or single-pixel imaging, for example) and more flexible in terms of hardware. We have shown elsewhere [40] that IGI can be implemented on a chip, which will greatly reduce the size and cost of an IGI system. These features can greatly accelerate the adoption of IGI for practical applications.

## Funding

## Disclosures

The authors declare no conflicts of interest.

## Reference

1. T. B. Pittman, Y. H. Shih, D. V. Strekalov, and A. V. Sergienko, "Optical imaging by means of two-photon quantum entanglement," Phys. Rev. A **52**, (1995).
2. R. S. Bennink, S. J. Bentley, and R. W. Boyd, ""Two-photon" coincidence imaging with a classical source," Phys. Rev. Lett. **89**, 113601 (2002).
3. F. Ferri, D. Magatti, A. Gatti, M. Bache, E. Brambilla, and L. A.Lugiato, "High-resolution ghost image and ghost diffraction experiments with thermal light," Phys. Rev. Lett. **94**, 183602 (2005).
4. A.Valencia, G. Scarcelli, M. D'Angelo, and Y. Shih, "Two-photon imaging with thermal light," Phys. Rev. Lett. **94**, 063601 (2005).
5. D. Z. Cao, J. Xiong, and K. G. Wang, "Geometrical optics in correlated imaging systems." Phys. Rev. A **71**, 013801 (2005).
6. L. Basano and P. Ottonello, "Experiment in lensless ghost imaging with thermal light," Appl. Phys. Lett. **89**, 091109 (2006).
7. Z. Yang, O.S. Magaña-Loaiza, M. Mirhosseini, Y. Zhou, B. Gao, L. Gao, S.M.H. Rafsanjani, G.L. Long, and R.W. Boyd, "Digital spiral object identification using random light," Light: Sci. & Appl. **6**(7), 17013 (2017).
8. J. Liu, J. Wang, H. Chen, H. Zheng, Y. Liu, Y. Zhou, F. L. Li and Z. Xu, "High visibility temporal ghost imaging with classical light," Opt. Commun. **410**, 824-829 (2018).
9. S. Y. Meng, Y. H. Sha, Q. Fu, Q. Q. Bao, W. W. Shi, G. D. Li, X. H. Chen, and L. A. Wu. "Super-resolution imaging by anticorrelation of optical intensities." Opt. Lett. **43**, 19 (2018).
10. Q. Fu, Y. Bai, X. Huang, S. Nan, P. Xie, and X. Fu, "Positive influence of the scattering medium on reflective ghost imaging," Photonics Res. **7**(12), 1468 (2019).
11. H. Wu, P. Ryczkowski, A. T. Friberg, J. M. Dudley, and G. Genty, "Temporal ghost imaging using wavelength conversion and two-color detection," *Optica* **6**(7), 902-906 (2019).
12. G. Barbastathis, A. Ozcan, and G. Situ, "On the use of deep learning for computational imaging," *Optica* **6**(8), 921-943 (2019).
13. Y. Kohno, K. Komatsu, R. Tang, Y. Ozeki, Y. Nakano, and T. Tanemura, "Ghost imaging using a large-scale silicon photonic phased array chip," Opt. Express **27**(3), 3817-3823 (2019).
14. Z. Ye, P. Qiu, H. Wang, J. Xiong, and K. Wang, "Image watermarking and fusion based on Fourier single-pixel imaging with weighed light source," Opt. Express **27**(25), (2019).
15. S. Sun, J. H. Gu, H. Z. Lin, L. Jiang, and W. T. Liu, "Gradual ghost imaging of moving objects by tracking based on cross correlation," Opt. Lett. **44**(22), (2019).
16. C. Hu, Z. Tong, Z. Liu, Z. Huang, J. Wang, and S. Han, "Optimization of light fields in ghost imaging using dictionary learning," Opt. Express **27**(20), (2019).
17. J. Wu, J. Wang, Y. Nie, and L. Hu, "Multiple-image optical encryption based on phase retrieval algorithm and fractional Talbot effect，" Opt. Express, 27(24), 35096-35107 (2019).
18. P. Clemente, V. Durán, E. Tajahuerce, and J. Lancis, "Optical encryption based on comaputational ghost imaging," Opt. Lett. **35**, (2010).
19. C. Zhao, W. Gong, M. Chen, E. Li, H. Wang, W. Xu, and S. Han, "Ghost imaging lidar via sparsity constraints," Appl. Phys. Lett. **101**, 141123 (2012).
20. W. Gong, C. Zhao, H. Yu, M. Chen, W. Xu, and S. Han, "Three-dimensional ghost imaging lidar via sparsity constraint," Sci. Rep. **6**, 26133 (2016).
21. D. Pelliccia, A. Rack, M. Scheel, V. Cantelli, and D. M. Paganin, "Experimental x-ray ghost imaging," Phys. Rev. Lett. **117**, 113902 (2016).
22. A. X. Zhang, Y. H. He, L. A. Wu, L. M. Chen, and B. B. Wang, "Tabletop x-ray ghost imaging with ultra-low radiation," Optica **5**, 374-377 (2018).
23. Z. Sun, F. Tuitje, and C. Spielmann, "Toward high contrast and high-resolution microscopic ghost imaging," Opt. Express **27**(5), (2019).
24. U. Rossman, R. Tenne, O. Solomon, I. Kaplan-Ashiri, T. Dadosh, Y. C. Eldar, and D. Oron, "Rapid quantum image scanning microscopy by joint sparse reconstruction," *Optica* **6**(10), 1290-1296 (2019).

25. Y. Qian, R. He, Q. Chen, G. Gu, F. Shi, and W. Zhang, "Adaptive compressed 3D ghost imaging based on the variation of surface normal," Opt. Express, **27**(20), 27862-27872 (2019).

26. K. Soltanlou and H. Latifi, "Three-dimensional imaging through scattering media using a single pixel detector," Appl. Optics, **58**(28), 7716-7726 (2019).

27. M. J. Sun, M. P. Edgar, G. M. Gibson, B. Sun, N. Radwell, R. Lamb, and M. J.Padgett. "Single-pixel three-dimensional imaging with time-based depth resolution," Nat. Comm. **7**(1) 12010 (2016).

28. B. Sun, M. P. Edgar, R. Bowman, L. E. Vittert, S. Welsh, A. Bowman, and M. J. Padgett. "3D computational imaging with single-pixel detectors," Science **340**, 844-847 (2013).

29. O. Katz, B. Yaron, and Y. Silberberg. "Compressive ghost imaging," Appl. Phys. Lett. **95**(13) 131110 (2009).

30. B. Yaron, O. Katz, and Y. Silberberg. "Ghost imaging with a single detector," Phys. Rev. A **79**(5) (2009): 053840.

31. Y. Wang, Y. Liu, J. Suo, G. Situ, C. Qiao, and Q. Dai, "High speed computational ghost imaging via spatial sweeping," Sci. Rep. **7,** 45325 (2017).

32. K. M. Czajkowski, A. Pastuszczak, and R. Kotyński. "Real-time single-pixel video imaging with Fourier domain regularization," Opt. Express **26**(16), 20009-20022 (2016).

33. Z. H. Xu, W. Chen, J. Penuelas, M. Padgett, and M. J. Sun. "1000 fps computational ghost imaging using LED-based structured illumination," Opt. Express **26**(3), 2427-2434 (2018).

34. J. Cheng. "Ghost imaging through turbulent atmosphere." Opt. Express **17**(10), 7916-7921 (2009).

35. R. E. Meyers, K. S. Deacon, and Y. Shih, "Turbulence-free ghost imaging," Appl. Phys. Lett. **98**(11), 111115 (2011).

36. M. Bina, D. Magatti, M. Molteni, A. Gatti, L. A. Lugiato, and F. Ferri. "Backscattering differential ghost imaging in turbid media," Phys. Rev. Lett. **110**(8), 083901 (2013).

37. M. Le, G. Wang, H. Zheng, J. Liu, Y. Zhou, and Z. Xu. "Underwater computational ghost imaging," Opt. Express, **25**(19), 22859-22868 (2017).

38. Z. Yang, L. Zhao, X. Zhao, W. Qin, and J. Li, "Lensless ghost imaging through the strongly scattering medium," Chin. Phys. B **25**(2), 024202 (2015).

39. C. Deng, L. Pan, C. Wang, X. Gao, W. Gong, and S. Han. "Performance analysis of ghost imaging lidar in background light environment," Photonics Res. **5**(5), 431-435 (2017).

40. Z. Yang, W. X. Zhang, Y. P. Liu, D. Ruan, and J. L. Li, "Instant Ghost Imaging: Algorithm and On-chip Implementation", arxiv: 1912.02181, (on the peer review process of Optics Express)

41. J. L. Li, "Study on second-order correlated imaging with pseudo-thermal light," PhD thesis. Tsinghua University, 2016 (In Chinese).

42. J. L. Li, Z. Yang, and L. L. Long, Patent CN201510151008.2 (2015) (In Chinese).

43. Z. Yang, J. L. Li and W. X. Zhang, Patent CN201910795456.4 (2019) (In Chinese).

44. H. H. Bossel, W. J. Hiller, and G. E. A. Meier. Noise-cancelling signal difference method for optical velocity measurements. Journal of Physics E: Scientific Instruments, 5(9), 893 (1972).

45. M. Coker, and D. Simkins (1980, April). A nonlinear adaptive noise canceller. In ICASSP'80. IEEE International Conference on Acoustics, Speech, and Signal Processing (Vol. 5, pp. 470-473). IEEE.